\documentclass[12pt]{article}

\usepackage{graphicx,url}

\usepackage[utf8]{inputenc}  
\usepackage{authblk} % for authors' affiliations

\usepackage{amsthm,amsmath}
\usepackage{natbib}
\usepackage{hyperref}
\usepackage{color}
    
\title{NEAT: an efficient network enrichment analysis test}

\author[1,2]{Mirko Signorelli\footnote{m.signorelli@rug.nl}}
\author[3]{Veronica Vinciotti}
\author[1]{Ernst C. Wit}

\affil[1]{Johann Bernoulli Institute, University of Groningen (NL)}
\affil[2]{Department of Statistical Sciences, University of Padova (IT)}
\affil[3]{Department of Mathematics, Brunel University London (UK)}

\date{}

\begin{document} 

\maketitle

\textcolor{red}{\textbf{ORIGINAL ARTICLE}\\
Please cite this paper as:\\
Signorelli, M., Vinciotti, V., Wit, E. C. (2016). NEAT: an efficient network enrichment analysis test. \textit{BMC Bioinformatics}, 17:352. DOI: 10.1186/s12859-016-1203-6.\\
The original version of this manuscript is freely accessible \hyperref{https://bmcbioinformatics.biomedcentral.com/articles/10.1186/s12859-016-1203-6}{}{}{from the website of BMC Bioinformatics}.}

\begin{abstract}
\noindent
Network enrichment analysis is a powerful method, which allows to integrate gene enrichment analysis with the information on relationships between genes that is provided by gene networks. Existing tests for network enrichment analysis deal only with undirected networks, they can be computationally slow and are based on normality assumptions.\\
We propose NEAT, a test for network enrichment analysis. The test is based on the hypergeometric distribution, which naturally arises as the null distribution in this context. NEAT can be applied not only to undirected, but to directed and partially directed networks as well. Our simulations indicate that NEAT is considerably faster than alternative resampling-based methods, and that its capacity to detect enrichments is at least as good as the one of alternative tests. We discuss applications of NEAT to network analyses in yeast by testing for enrichment of the Environmental Stress Response target gene set with GO Slim and KEGG functional gene sets, and also by inspecting associations between functional sets themselves.\\
NEAT is a flexible and efficient test for network enrichment analysis that aims to overcome some limitations of existing resampling-based tests. The method is implemented in the \texttt{R} package \texttt{neat}, which can be freely downloaded from CRAN (\hyperref{https://cran.r-project.org/package=neat}{}{}{https://cran.r-project.org/package=neat}).
\hspace{0.3cm}\\
\noindent \textbf{Keywords:} network; enrichment analysis; gene expression; hypergeometric.
\end{abstract}

\section{Background}
The advent of high throughput technologies has driven the development of cell biology over the last decades. The diffusion of microarrays and next generation sequencing techniques has made available a large amount of data that can be used to increase our understanding of gene expression. The need to analyse and interpret these data has led to the development of new methods to infer relationships between genes, which require a combination of biological knowledge, statistical modelling and computational techniques.\\
When the first data on gene expression became available, they were usually analysed considering each gene separately. However, researchers soon realized that genes act in a concerted manner, and that cellular processes are the result of complex interactions between different genes and molecules. Nowadays, sets of genes that are responsible for many cellular functions have been identified, and are collected in publicly available databases \citep{ashburner2000,kanehisa2000}.\\
One of the advantages of these sets of genes, whose function is already known, is that they can be used to interpret the results of new experiments: this has led to the implementation of a large number of methods for \textit{gene enrichment analysis} \citep{huang2009}. Their aim is to compare gene expression levels under two different conditions (experimental vs control), and to detect which sets of genes are differentially expressed (enriched) in the experimental condition. To this end, genes are ordered in a list $L$ in decreasing order of differential expression, and enrichment is then tested in different ways. \textit{Singular enrichment analysis} \citep{robinson2002,beissbarth2004} tests the over or under-representation of functional gene sets within the set of genes defined by the first $k$ top genes in $L$. The major limitations of this approach lie in the fact that the choice of $k$ is arbitrary, and that the test does not take into account gene expression levels. \textit{Gene set enrichment analysis} \citep{subramanian2005,kim2005} overcomes these limitations, by making use of the whole list $L$ of genes, and testing the tendency of genes belonging to a functional set to occupy positions at the top (or at the bottom) of $L$. A limitation that is common to both single and gene set enrichment analysis, however, is that these methods base computations on the level of overlap between sets of genes only, without considering associations and interactions between genes.\\
Gene networks are an established tool to represent these interactions. In \textit{network inference} \citep{desmet2010,marbach2010}, genes or molecules are represented as nodes of a graph and their interactions are modelled as links between the nodes. These links can be represented as either a directed or an undirected edge, and a graph is called directed if all edges are directed, undirected if every edge is undirected and partially directed (or mixed) otherwise \citep{Lauritzen1996}. An undirected edge displays association between two genes, while a directed edge posits a direction in the relationship between them. Network estimation represents a difficult task, and many different estimation methods have been proposed \citep{friedman2008,abegaz2013}. \cite{marbach2012} classified them into six groups and pointed out that their predictive performance can vary a lot within each group and according to the structure of the network. In order to integrate evidence on gene associations unveiled by a number of experimental and computational studies into a single network, curated gene networks for different species have been proposed, including \textit{YeastNet} \citep{kim2013} and \textit{FunCoup} \citep{schmitt2014}.\\
In an attempt to integrate the information on interactions between genes provided by gene networks into enrichment analyses, researchers have recently developed methods for \textit{network enrichment analysis} \citep{shojaie2010,glaab2012,alexeyenko2012,mccormack2013}. The idea, here, is to test enrichment between sets of genes in a network. \cite{shojaie2010} focus mainly on network inference, proposing to represent the gene network with a linear mixed model, so that enrichment tests can be then computed by testing a system of linear hypotheses on the fixed effect parameters of the model.
\cite{glaab2012}, \cite{alexeyenko2012} and \cite{mccormack2013}, instead, assume that a gene network is already available (either from the literature or as the result of a tailored inferential process) and focus their attention on the strategy that can be used to assess enrichment between sets of nodes. In particular, \cite{glaab2012} propose a network enrichment score based on a suitably defined network distance between two sets of nodes, alongside an empirical method for setting a cut-off on this distance. In contrast to this, \cite{alexeyenko2012} and \cite{mccormack2013} derive network enrichment scores on the basis of statistical tests against the null distribution of no enrichment. The advantage of the approach proposed by Alexeyenko et al. and McCormack et al. is that the assessment of enrichment is based on a significance testing procedure.\\
The idea of \cite{alexeyenko2012} and \cite{mccormack2013} is that the presence of enrichment between two sets of genes, say $A$ and $B$, can be assessed by comparing the number of links connecting nodes in $A$ and $B$ with a reference distribution, which models the number of links between the same two sets in the absence of enrichment. Both \cite{alexeyenko2012} and \cite{mccormack2013} assume that the reference distribution is approximately normal, and they obtain its mean and variance by means of permutations, i.e., computing the mean and variance of the number of links between $A$ and $B$ in a sequence of random replications of the network. Their tests rely on algorithms that permute the network, and mainly differ between themselves for the fact that each algorithm aims to preserve different topological properties of the original network in the generation of network replicates. These methods, however, suffer from three limitations. First of all, they require the simulation of a large number of permuted networks, an activity that can be computationally intensive and highly time consuming (especially for big networks). Furthermore, they base the computation of the test on a normal approximation for the reference distribution, whose nature is discrete. \cite{mccormack2013} show that such an approximation is inaccurate when the expected number of links between $A$ and $B$ is small. A further drawback of these methods is that they have been implemented so far only for undirected networks.\\
In this work we build upon the approach of \cite{alexeyenko2012} and \cite{mccormack2013} and propose an alternative test which we call NEAT (Network Enrichment Analysis Test). The main idea behind this test is that, under the null hypothesis of no enrichment, the number of links between two gene sets $A$ and $B$ follows an hypergeometric distribution. This enables us to model the reference distribution directly via a discrete distribution, without having to resort to a normal approximation. NEAT does not require network permutations to compute mean and variance under the null hypothesis, and is therefore faster than the existing resampling-based methods. Moreover, we develop NEAT not only for undirected, but also for directed and partially directed networks, thus providing a common framework for the analysis of different types of networks.

\section{Methods}
\begin{figure}[tb]
\begin{center}
	\includegraphics[scale=0.6, trim = {2.7cm 3cm 1cm 3cm}]{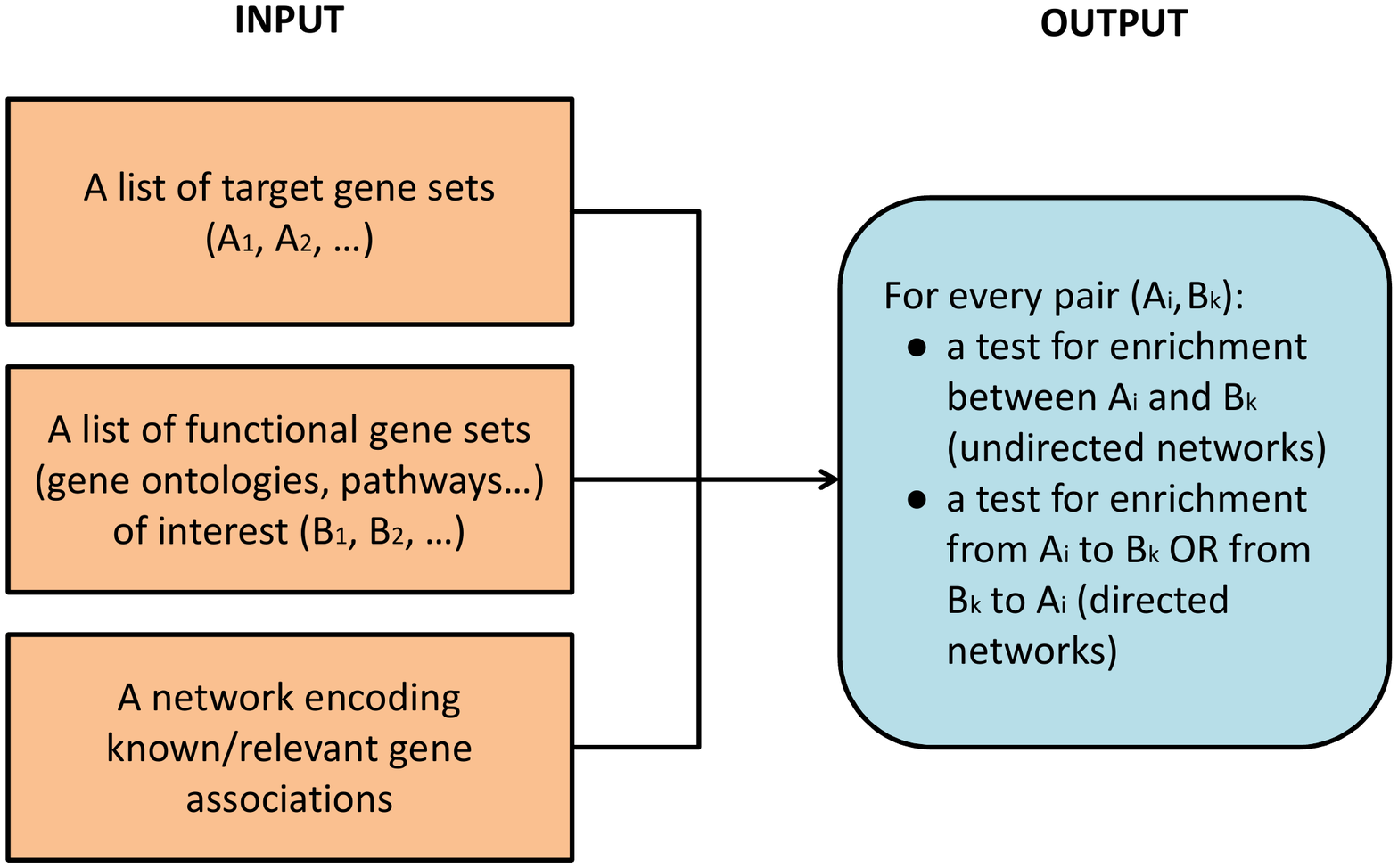} %left, low, right, upp
\end{center}
	\caption{\textit{Workflow diagram of a typical network enrichment analysis with NEAT.}
	}\label{fig:workflowdiagr}
      \end{figure}

The starting point of enrichment analyses is the identification of one or more gene sets of interest. These target gene sets are typically groups of genes that are differentially expressed between experimental conditions, but they can also be different types of gene sets: e.g., clusters of genes that are functionally similar in a given time course, or genes that are bound by a particular protein in a ChIP-chip or ChIP-seq experiment. Enrichment analysis provides a characterization of each target gene set by testing whether some known functional gene sets can be related to it. Methods for gene enrichment analysis assess the relationship between a target gene set and each functional gene set simply by considering the overlap of these two groups. In contrast to this, network enrichment analysis incorporates  an evaluation of the level of association between genes in the target set and genes in the functional gene set into the test.\\
Information on associations and dependences between genes is represented by a network, which consists of a set of $N$ nodes $V = \{v_1,...,v_N\}$ that are connected by edges (links). Each gene is thus represented as a node $v_i$ of the network, and a link between two nodes is drawn to signify interaction between the corresponding genes. Examples of genome-wide curated networks that collect known gene associations are \textit{YeastNet} \citep{kim2013} and \textit{FunCoup} \citep{schmitt2014}.\\
A natural way to study the relation between two sets of genes $A$ and $B$ in a network is to consider the presence or absence of links connecting nodes in the two groups \citep{alexeyenko2012,mccormack2013}. In the inferred network, we expect that individual links may be slightly unstable and noisy. However, we do expect that the inferred links contain a sign of the relationships between gene sets. So, although links between individual genes in sets $A$ and $B$ may be noisy, if there is a functional relationship between functions described by sets $A$ and $B$ we expect the number of links between the two groups to be larger (or smaller) than expected by chance. If this is the case, we say that there is enrichment between $A$ and $B$.\\
Links between two nodes of a network can be either directed (arrows) or undirected. The presence of an arrow between two genes implies a directionality in the relation between them, whereas an undirected edge does not provide information on the direction of the relation. The upcoming subsection considers directed networks. In this case, one can distinguish two cases: whether genes in the target set regulate genes of the functional set, or genes in the functional gene set regulate genes in the target set (enrichment from $A$ to $B$, or from $B$ to $A$). This distinction does not occur for undirected networks, which are the subject of the next subsection: in this case, $A$ and $B$ are exchangeable, and we simply talk of enrichment ``between'' $A$ and $B$. A workflow diagram summarizing the input and the output of NEAT is shown in Figure \ref{fig:workflowdiagr}.

\subsection{Enrichment test for directed networks}
\begin{figure}[tb]
\begin{center}
	\includegraphics[scale=0.40]{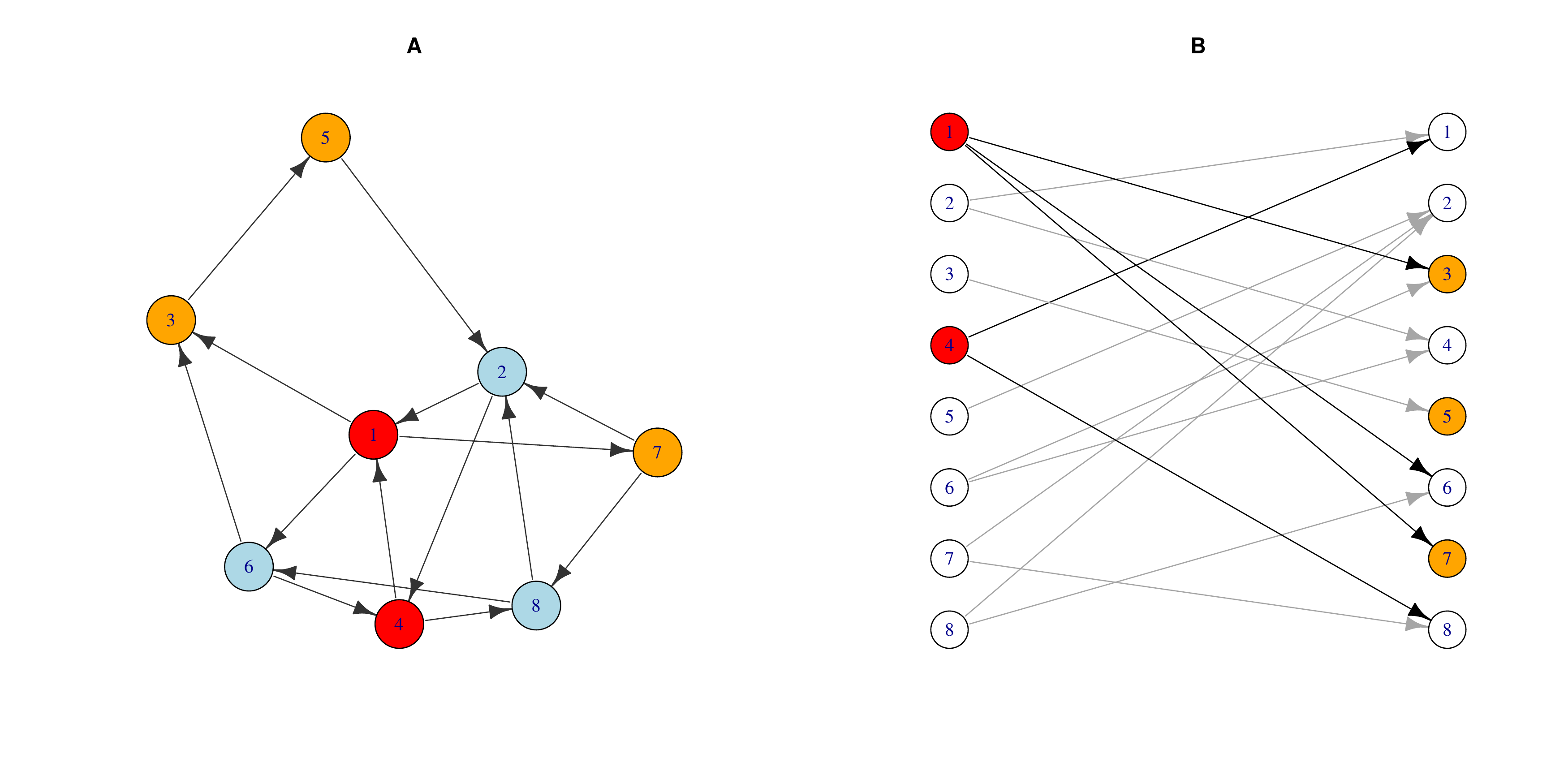}
\end{center}
  \caption{\textbf{Example: NEAT in directed networks.}
      \textit{Left:} directed network consisting of 8 nodes connected by 15 arrows. Set $A$ contains nodes 1 and 4 (red) and set $B$ nodes 3, 5 and 7 (orange).
      \textit{Right:} bipartite representation of the same network: it can be observed that $n_{AB}=2$, $o_A=5$, $i_B=4$ and $i_V=15$. It follows that $\mu_0=1.07$ and $p=0.48$.}\label{fig:example}
      \end{figure}

In a directed network, we assess the presence of enrichment from $A$ to $B$ by considering the number of arrows going from genes in $A$ to genes belonging to $B$. We denote this by $n_{AB}$. The observed $n_{AB}$ can be thought of as a realization from a random variable $N_{AB}$, with expected value $\mu_{AB}$. To assess the relation from $A$ to $B$, we compare $\mu_{AB}$ with the number of arrows that we would expect to observe from $A$ to $B$ by chance, which we denote as $\mu_0$. We say that there is enrichment from $A$ to $B$ if $\mu_{AB}$ is different from $\mu_0$. Furthermore, we say that there is over-enrichment from $A$ to $B$ if $\mu_{AB}$ is higher than $\mu_0$, and under-enrichment (or depletion) if $\mu_{AB}$ is lower than $\mu_0$.\\
We propose a test based on the hypergeometric distribution to assess the significance of this difference. The motivation behind this choice is the following. The hypergeometric distribution models the number of successes in a random sample without replacement: in our case, we can mark arrows in the network that reach genes in $B$ as ``successful'', and the remaining ones as ``unsuccessful''. Then, we can view the arrows that go out from genes in $A$ as a random sample without replacement from the population of arrows present in the graph: if there is no relation (i.e., no enrichment) between $A$ and $B$, then the distribution of $N_{AB}$ (the number of successes in the sample) is
\begin{equation}
	N_{AB} \sim \mbox{hypergeom}(n = o_A, K = i_B, N = i_V), \label{dirassumption}
\end{equation}
where the sample size $o_A$ is the outdegree of $A$ (the total number of arrows going out from genes that belong to $A$), the number of successful cases in the population $i_B$ is the indegree (number of incoming arrows) of $B$ and the population size $i_V$ is the total indegree of the network (which is equal to the total number of arrows).\\
It is certainly possible to imagine alternative choices for the null distribution of $N_{AB}$. \cite{alexeyenko2012} and \cite{mccormack2013} assume that $N_{AB}$ is normal with mean $\mu_0$ and variance $\sigma^2_0$, and they use network permutations to estimate $\mu_0$ and $\sigma^2_0$. However, the normal distribution is continuous and symmetric, so that their choice implies somehow that the behaviour of $N_{AB}$ should be roughly symmetric, and could be well approximated with a continuous random variable. In addition, estimation of $\mu_0$ and $\sigma^2_0$ by means of network permutations can be highly time consuming. Alternatively, one could consider for $N_{AB}$ an hypergeometric distribution with different parameters, defined for example, by considering all possible edges in the network (instead of the edges that are actually present in the network) as a population. We prefer model (\ref{dirassumption}) over this alternative, because the choice of the parameters therein allows to condition on two quantities that we consider crucial, which are the outdegree of $A$ and the indegree of $B$. Moreover, in our experience so far, we have observed that tests based on alternative parametrizations often result in poor performances.\\
The null mean and variance of $N_{AB}$ can be immediately derived from model (\ref{dirassumption}). In particular, in the absence of enrichment we expect to observe, on average, $\mu_0 = o_A \frac{i_B}{i_V}$ arrows from nodes in $A$ to nodes in $B$. Thus, we expect $\mu_0$ to increase as the number of arrows leaving $A$, or reaching $B$, increases.
Biological assessment of enrichment can therefore be carried out by testing the null hypothesis of no enrichment
$$ H_0: \mu_{AB} = \mu_0 $$
against the alternative hypothesis of enrichment
$$ H_1: \mu_{AB} \neq \mu_0.$$
In a test with a discrete test statistic and two-sided alternative, such as the one that we propose, the p-value can be computed in different ways \citep{gibbons1975,blaker2000,agresti2013}. Let $T$ be a discrete test statistic and $t$ be the observed value of $T$. A first possibility is to compute the p-value for the two-tailed test by doubling the one-tailed p-value, $p_1 = 2 \min [ P_0(T \leq t), P_0(T \geq t)]$, where $P_0$ denotes the distribution of $T$ under the null hypothesis. An evident drawback of this formula, however, is that $p_1$ can exceed 1, and therefore $p_1$ does not represent a probability. Even though a simple modification $p_2 = \min(p_1,1)$ could avoid the problem, we prefer to subtract $P_0(T=t)$ from $p_1$ ($P_0(T=t)$ is non-null for discrete $T$, and this is the reason why $p_1$ can exceed 1) and to compute the p-value using
\begin{align}
	p &= 2 \min [ P_0(T < t), P_0(T > t)] + P_0( T=t) \label{pvalue}\\
	& = 2 \min \left[ P_0(N_{AB}>n_{AB}), P_0(N_{AB}<n_{AB}) \right] + P_0(N_{AB} = n_{AB}),\nonumber
\end{align}
which always lies within the interval $[0,1]$ and differs from $p_1$ by a factor equal to $P_0(T=t)$. A $p$-value close to 0 can be regarded as evidence of enrichment, because it entails that the number of links from $A$ to $B$ is significantly smaller or higher than we would expect it to be in the absence of enrichment. Therefore, for a given type I error probability $\alpha$, we conclude that there is evidence of enrichment from $A$ to $B$ if $p < \alpha$, while if $p \geq \alpha$ there is not enough evidence of enrichment.\\
As an example, consider the network in Figure \ref{fig:example}. Suppose that we are interested to test whether there is enrichment from the set $A=\{1,4\}$ to the set $B=\{3,5,7\}$. It can be observed that there are 5 arrows going out from $A$, and 2 of them reach $B$. The whole network consists of 15 arrows, of which 4 reach $B$. Thus, $n_{AB} = 2$, $o_A = 5$, $i_B=4$ and $i_V=15$. The idea behind (\ref{dirassumption}) is that, if the 5 arrows that are going out from $A$ are a random sample (without replacement) from the 15 arrows that are present in the network, then the proportion of arrows reaching $B$ from $A$ should be close to the proportion of arrows reaching $B$ in the whole network, and in the absence of enrichment we should observe on average $\mu_0 = 1.33$ edges. In this case, it seems that arrows going out from $A$ tend to reach $B$ more frequently ($40\%$) than other arrows do ($27\%$ of the $15$ arrows in the network reach $B$). However, the computation of the p-value leads to $p = 0.48$: the observed $n_{AB} = 2$ does not provide enough evidence to reject the null hypothesis, so that the conclusion of the test is that there is no enrichment from $A$ to $B$.

We can also consider sets $B=\{3,5,7\}$ and $C=\{2,5\}$ (note that the two groups share gene 5), and test enrichment from $B$ to $C$. In this case, $n_{BC} = 3$ arrows out of $o_B=4$ (75\%) reach $C$ from $B$, whereas in the whole network $i_C=4$ arrows out of $d_V = 15$ (27\%) reach $C$. The null expectation is here $\mu_0 = 1.07$; if we fix the type I error probability equal to $\alpha = 5\%$, the p-value $p = 0.03$ leads to the conclusion that there is enrichment from $B$ to $C$.

\subsection{Enrichment test for undirected networks}
\begin{figure}[tb]
\begin{center}
\includegraphics[scale=0.4]{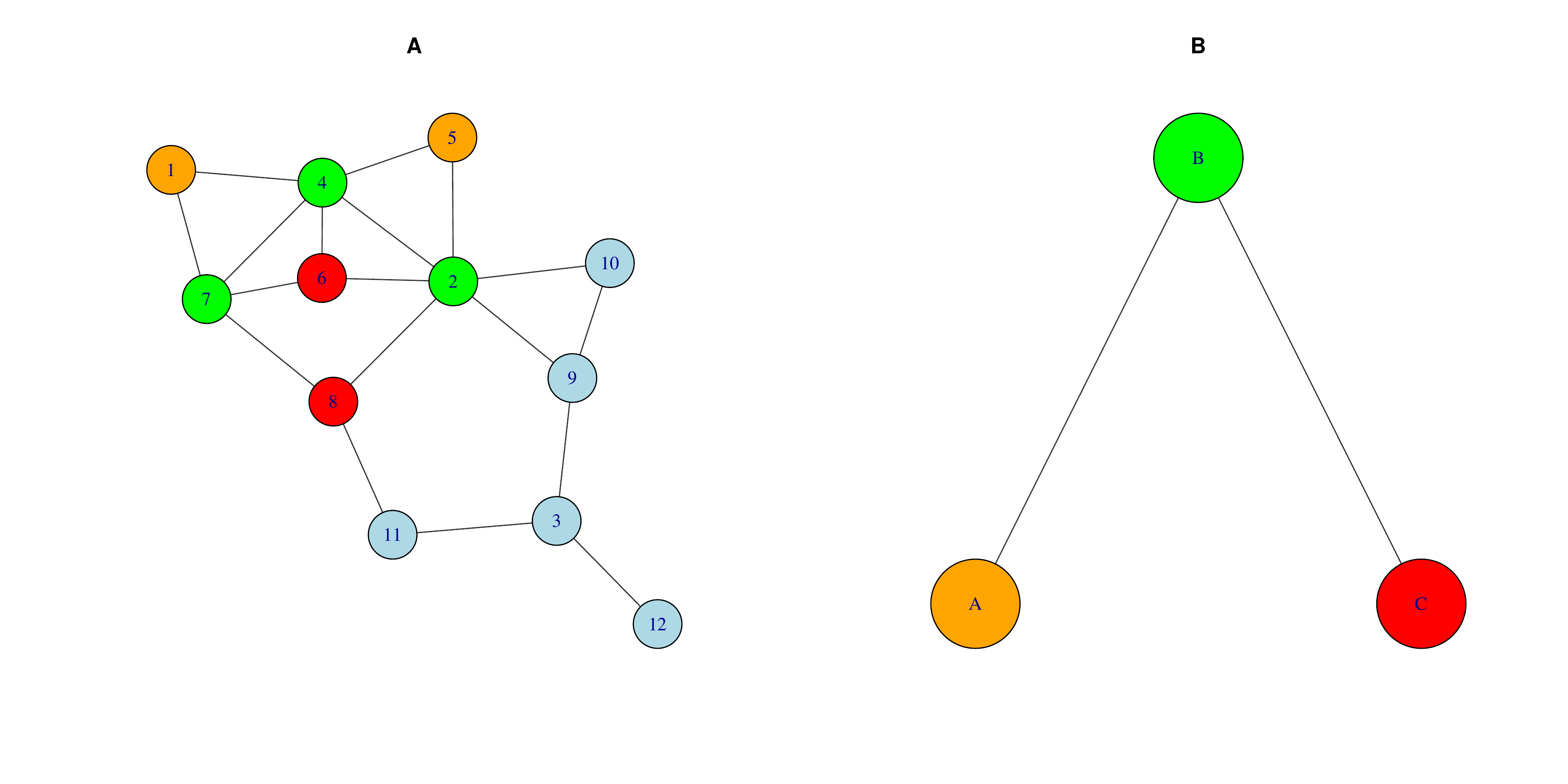}
\end{center}	
  \caption{\textbf{Example: NEAT in undirected networks.}
  	\textit{Left:} undirected network with 12 nodes. We are interested to infer the relation between sets A (nodes 1 and 5), B (2, 4 and 7) and C (6 and 8).
  	\textit{Right:} representation of the relations between sets: enrichment is detected between sets A and B ($p = 0.023 $) and between sets B and C ($p = 0.038$), but not between sets A and C ($p = 0.465$).}\label{fig:example2}
      \end{figure}

When dealing with undirected networks, the presence of enrichment between $A$ and $B$ is assessed considering the number of edges that connect genes in $A$ to genes in $B$. We denote this by $n_{AB}$. Given the undirected nature of the links in the network, there is no distinction between indegree and outdegree of a node, and it only makes sense to consider the degree of a node, which is the number of vertices that are linked to that node. The null distribution (\ref{dirassumption}) should thus be adapted accordingly. Let us define the total degree $d_S$ of a set $S$ as the sum of the degrees of nodes that belong to it: then, in the absence of enrichment we can view $n_{AB}$ as the number of successes in a random sample of size $d_A$, drawn from a population of size $d_V$. The null distribution of $N_{AB}$ for undirected networks is thus
\begin{equation*}
	N_{AB} \sim \mbox{hypergeom}(n = d_A, K = d_B, N = d_V), %\label{undirassumption}
\end{equation*}
where $d_A$, $d_B$ and $d_V$ are the total degrees of sets $A, B$ and $V$.\\
The null hypothesis is then that $\mu_{AB} = \mu_0 = d_A \frac{d_B}{d_V}$, the alternative that $\mu_{AB} \neq \mu_0$. The p-value is computed using formula (\ref{pvalue}).\\
As an example, consider the network in Figure \ref{fig:example2}A and suppose that we are interested to test the presence of enrichment between the pairs of sets $(A,B)$, $(A,C)$ and $(B,C)$. Sets $A$ and $B$ are linked by $n_{AB}=4$ edges, and their degrees are $d_A=4$ and $d_B=15$, while $d_V= 36$. Thus, $\mu_0 = 1.67$ and $p^{AB}= 0.023$. In the same way, it is possible to compute $p^{AC}=0.465$ and $p^{BC}=0.038$. Figure \ref{fig:example2}B shows the relation between the three sets fixing $\alpha = 5\%$: enrichment is present between the pairs $(A,B)$ and $(B,C)$, but not between sets $A$ and $C$.

\subsection{Enrichment test for partially directed networks}
A partially directed network (or ``mixed'' network) is a network where both directed and undirected edges are present. It is possible to view such a network as a directed network, where every undirected edge connecting two nodes $v$ and $w$ represents in fact a pair of arrows, the former going from $v$ to $w$ and the latter from $w$ to $v$. If such an adaptation is adopted, model (\ref{dirassumption}) can be applied and partially directed networks can be analysed within \texttt{neat} as directed networks.

\subsection{Software}
NEAT is implemented in the \texttt{R} package \texttt{neat} \citep{neat-cran}, which can be freely downloaded from CRAN: \hyperref{https://cran.r-project.org/package=neat}{}{}{https://cran.r-project.org/package=neat}. The manual and a vignette illustrating the package are also available from the same URL. The package allows users to specify the network in different formats, it includes functions to plot and summarize the results of the analysis and is accompanied by a set of data and examples, including the enrichment analysis of the ESR gene sets that we discuss in Section \ref{sec:yeast}.

\section{Performance evaluation}
We assess the performance of NEAT by means of simulations. Table \ref{Tab:sima1} summarizes some aspects of these simulations, that are the subject of the next two subsections. The \texttt{R} scripts and data files for each simulation can be found at \hyperref{https://github.com/m-signo/neat}{}{}{https://github.com/m-signo/neat}.

\begin{table}[tb]
\caption{ \textbf{An overview of simulations S1-S5.} In Simulations S1 and S2, we compare the performance of NEAT in two directed networks with different degree distribution. In simulation S3, we check the performance of the test for different levels of overlap, ranging from $0\%$ to  $100\%$. In Simulations S4 and S5, we compare NEAT to alternative tests in two undirected networks with different degree distribution.
}\label{Tab:sima1}
\small 	\vspace{0.4cm} \centering
      \begin{tabular}{cccccc}
        \hline
        & & & & \multicolumn{2}{c}{Overlap:}\\
Simulation & Network type & Degree distribution	& Graph density & mean 	& maximum\\ \hline
S1 & Directed	& Power law	& 3\%	& 4\%	   	& 11.3\%\\
S2 & Directed	& Mixture of 2 Poisson	& 4\%	& 3.6\%		& 9.5\%\\
S3 & Directed	& Mixture of 2 Poisson	& 4\%	& - & -\\
S4 & Undirected	& Power law & 3\%	& 3.8\%		& 12\%\\
S5 & Undirected & Mixture of 2 Poisson	& 4\%	& 3.6\%		& 11\%\\\hline
      \end{tabular}
\end{table}

We first consider directed networks, and check whether the performance of NEAT is influenced by the degree distribution of the network, or by the level of overlap between sets of nodes. We then consider undirected networks, and carry out a comparison of NEAT with the NEA test of \cite{alexeyenko2012} and with the LP, LA, LA+S and NP tests of \cite{mccormack2013}.\\
We compare the performance of the methods under the null hypothesis by checking whether the empirical distribution of p-values in the absence of enrichment is uniform using the Kolmogorov-Smirnov test, and by computing the following ratios:
$$ R_1 = \frac{\text{Number of enrichments at } 1\% \text{ level}}{0.01 \times \text{Number of tests where } H_0 \text{ is true}} $$
and
$$R_5 = \frac{\text{Number of enrichments at } 5\% \text{ level}}{0.05 \times \text{Number of tests where } H_0 \text{ is true}}.$$
The idea behind $R_1$ and $R_5$ is that if the null hypothesis $H_0$ is true, we expect a good test to reject it with a frequency that is close to $\alpha$. So, the target value for $R_1$ and $R_5$ is 1.\\
Furthermore, we compare the capacity of different tests to correctly detect enrichments and non-enrichments by computing specificity and sensitivity at $\alpha = 5\%$ level, and the area under the ROC curve (AUC).
The specificity is the proportion of correctly detected non-enrichments, and we expect it to be as close as possible to $1-\alpha$. The sensitivity indicates the proportion of correctly detected enrichments, whereas the AUC is a measure of the overall capacity of a test to discriminate enrichments and non-enrichments across all values of $\alpha$. Therefore, a test will show a good performance whenever it achieves a specificity close to $1 - \alpha$, and values of sensitivity and AUC as high as possible (ideally 1).

\subsection{Simulation with directed networks}
In simulations S1 and S2, we generate two random networks with 1000 nodes and with fixed indegree and outdegree distributions using the algorithm implemented by \cite{csardi2006}. The indegree and outdegree distributions of nodes are power law with exponent $4$ and minimum degree 20 in simulation S1, and a mixture of two Poisson distributions, with parameters $\lambda_1 = 40$ and $\lambda_2 = 100$ and weights $q_1 = 99\%$ and $q_2 = 1\%$, in simulation S2.\\
We consider 50 sets of nodes whose size ranges between $50$ and $100$, and we test enrichment from $A$ to $B$ and from $B$ to $A$ for every pair of sets: this means that, in total, we compute $50 \times 49 = 2450$ tests. In the original networks, no preferential attachment (i.e., no enrichment) between any couple of these sets is present; we generate enrichments by increasing or reducing the number of arrows for 200 pairs of sets. In each case, enrichment is created by adding or removing arrows randomly from one group to the other, in such a way that $n_{AB}$ increases or reduces by a proportion uniformly ranging from $10\%$ to $50\%$.\\
Table \ref{Tab:sima2} shows that the empirical distribution of p-values in absence of enrichment is approximately uniform both in simulation S1 and S2. The sensitivity is higher in simulation S2, whereas the specificity is close to the target value (95\%) in both cases. As a result, the area under the ROC curve is slightly higher in simulation S2. Overall, the test shows in both cases a good capacity to discriminate enrichments and non-enrichments.

\begin{table}[tb]
	\caption{\textbf{Performance of NEAT in simulations S1 and S2.} $p^{KS}$ denotes the p-value of the Kolmogorov-Smirnov test for uniform distribution, AUC is an abbreviation for ``area under the ROC curve''. In both simulations, the distribution of p-values under $H_0$ is uniform and the specificity is close to the expected  $95 \%$ value. Sensitivity and AUC are higher in simulation S2.
}\label{Tab:sima2}
	\vspace{0.4cm} \centering
	\begin{tabular}{ccccccc}
		\hline
Simulation	& $p^{KS}$ & $R_1$ & $R_5$ & Sensitivity & Specificity & AUC\\ \hline
S1	& 0.510 & 1.56 & 1.17 & 73\% & 94\% & 0.894\\
S2	& 0.125 & 1.20 & 1.12 & 78\% & 94\% & 0.927\\\hline
	\end{tabular}
\end{table}

\begin{figure}[tb]
\centering
	\includegraphics[scale=0.6]{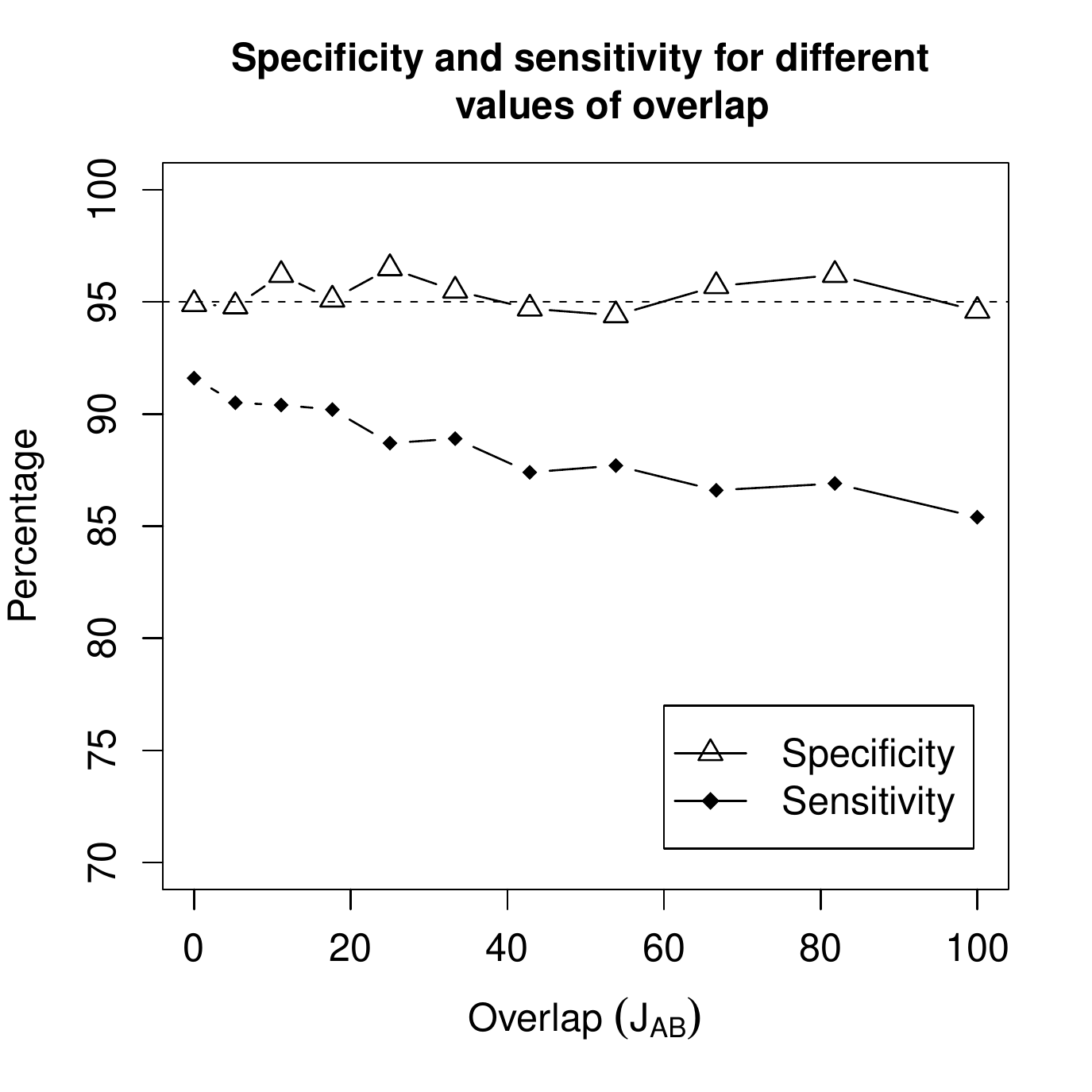}
	\caption{\textbf{Specificity and sensitivity in simulation S3.}
		The plot shows the values of specificity and sensitivity for different levels of overlap (every point in the plot is computed on the basis of 1000 tests). We observe that the specificity of the test does not vary substantially for different levels of overlap, and is always close to $95 \%$ as expected. The sensitivity, instead, slightly reduces as the percentage of overlap increases.
		}\label{fig:spesen}
\end{figure}

In simulation S3 we check whether the proportion of overlap between sets $A$ and $B$, that we measure with the Jaccard index
$$J_{AB} = |A \cap B| / |A \cup B|,$$
could have an effect on specificity and sensitivity. We consider the same network used in simulation S2, and we test enrichment between pairs of sets with fixed size $|A|=|B|=50$, but with increasing overlap (we consider $|A \cap B| \in \{0,5,10,15,...,50\}$). Under $H_0$ we do not modify the network, whereas under $H_1$ we introduce enrichments adding $35$ arrows going from genes in $A$ to genes in $B$. For every value of overlap, we consider 2000 test ($H_0$ is true in 1000 cases, and false in the remaining 1000). Figure \ref{fig:spesen} shows that the specificity remains constant and close to $95\%$ for any level of overlap; the sensitivity, on the other hand, is slightly higher when the level of overlap is moderate.

\subsection{Simulation with undirected networks}
As alternative methods for network enrichment analysis are available for undirected networks only, we compare NEAT with them in two simulations where we consider undirected networks with $1000$ nodes. We generate two random networks with fixed degree distribution, using the algorithm implemented by \cite{csardi2006}; the degree distribution follows a power law in simulation S4 and a mixture of Poisson distributions in simulation S5, with the same parameters used in simulations S1 and S2. Likewise, we consider 50 sets of nodes, whose sizes vary between 50 and 100 nodes. We test enrichment between every pair of sets $A$ and $B$, so that the total number of comparisons is here $50 \times 49 /2 = 1225$. We introduce enrichments for 100 pairs of sets by adding or removing edges randomly between them, in such a way that $n_{AB}$ is increased or reduced by a proportion uniformly ranging from $10\%$ to $50\%$.

\begin{table}[tb]
	\caption{\textbf{Results of simulation S4.} The best results for each indicator are in \textbf{bold}. $p^{KS}$ denotes the p-value of the Kolmogorov-Smirnov test for uniform distribution, AUC is an abbreviation for ``area under the ROC curve''. The distribution of p-values under $H_0$ is evidently not uniform for NEA and LP. NEAT shows the highest values of sensitivity and AUC, and its specificity is close to the target value (95\%).
}\label{Tab:simc}
	\vspace{0.4cm} \centering
	\begin{tabular}{ccccccc}
		\hline
Test & $p^{KS}$ & $R_1$ & $R_5$ & Sensitivity & Specificity & AUC\\ \hline
NEAT	& \textbf{0.399} & 1.33 		& \textbf{1.14} & \textbf{69}\% & \textbf{94}\% 	& \textbf{0.920}\\
NEA		& 0.001 		 & 0    		& \textbf{0.87} & \textbf{68}\% & \textbf{96}\% 	& \textbf{0.918}\\
LP		& 0				 & 2.13 		& 1.51 			& \textbf{68}\% & 92\% 				& 0.908\\
LA		& \textbf{0.255} & 1.60 		& 1.17 			& 60\% 		& \textbf{94}\% 	& 0.897\\
LA+S	& \textbf{0.409} & 1.87 		& 1.17 			& 63\% 		& \textbf{94}\% 	& 0.913\\
NP		& 0.037			 & \textbf{1.24} & 1.28 		&	58\% 	& \textbf{94}\% 	& 0.884\\ \hline
	\end{tabular}
\end{table}

\begin{figure}[tb]
\centering
	\includegraphics[scale=0.6]{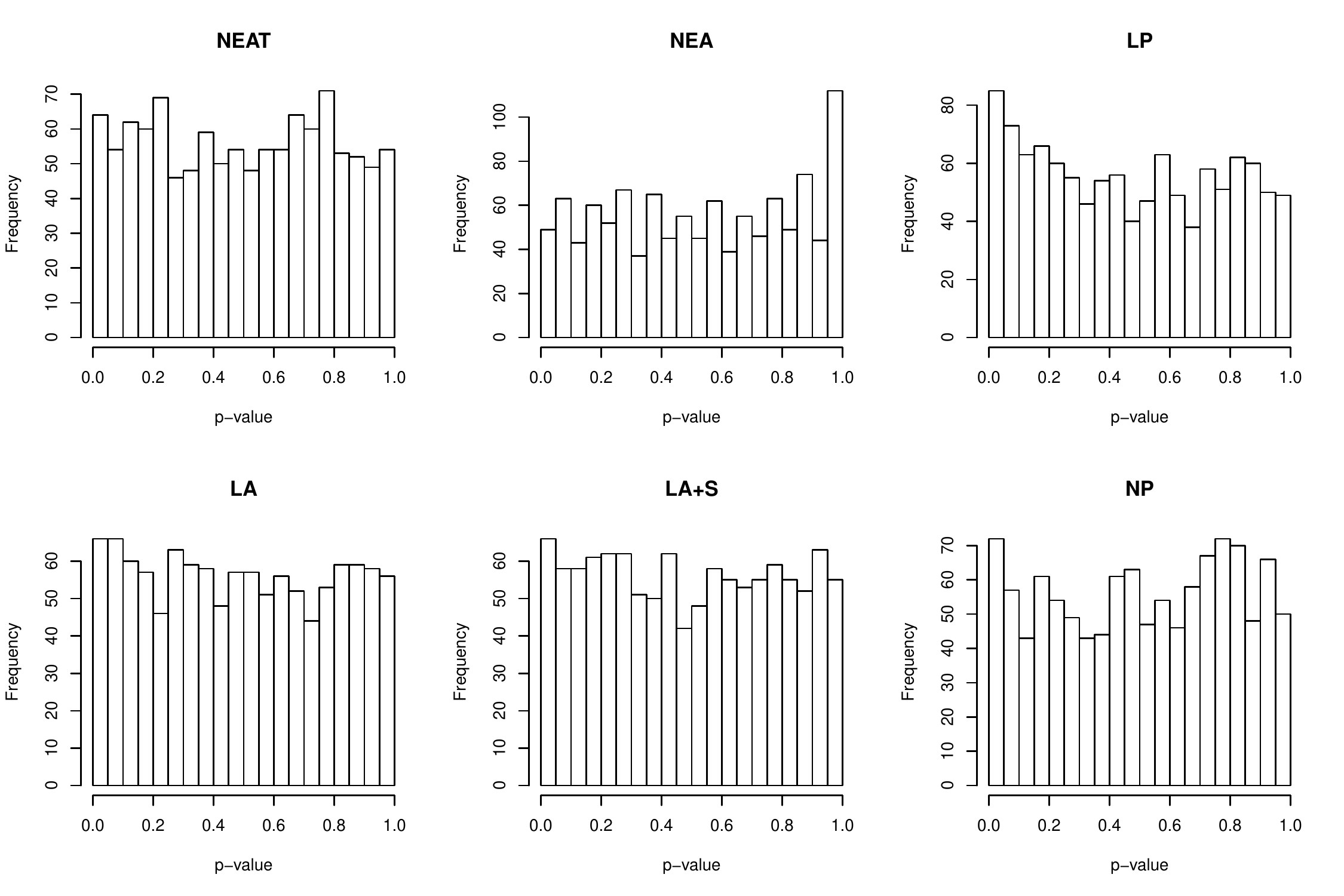}
	\caption{\textbf{Histogram of p-values in absence of enrichment in simulation S4.}
		 The test of Kolmogorov-Smirnov indicates that the distribution is uniform for NEAT ($p=0.34$), LA ($p=0.11$) and NP ($p=0.32$). The distribution of p-values is highly left-skewed for NEA, and right-skewed for LP.
		 }\label{fig:powlawhist}
\end{figure}

Tables \ref{Tab:simc} and \ref{Tab:simd} show the results for simulations S4 and S5, respectively. As concerns the behaviour under the null hypothesis, the distribution of p-values is uniform in both cases for NEAT and LA, and in one case for LA+S (simulation S4) and NP (S5). NEA and LP, instead, do not produce uniform distributions: as it can be observed from Figure \ref{fig:powlawhist}, the reason is that the distribution is strongly left-skewed for NEA, whereas for LP the distribution is right-skewed (the same patterns occur also in simulation S5). In both simulations, most of the methods achieve a specificity close to $95\%$ as expected; comparison with the other tests shows that the sensitivity and AUC of NEAT are overall good.\\
Table \ref{Tab:speed_sim} compares the speed of computation for the different methods. NEAT turns out to be the fastest method by far, being 22 times faster than NP (the fastest alternative) and more than 3000 times faster than NEA (the slowest alternative). This result is mostly due to the fact that NEAT does not require the generation of a large number of permuted networks to compute the test.

\begin{table}[tb]
	\caption{\textbf{Results of simulation S5.} The best results for each indicator are in \textbf{bold}. $p^{KS}$ denotes the p-value of the Kolmogorov-Smirnov test for uniform distribution, AUC is an abbreviation for ``area under the ROC curve''. The distribution of p-values under $H_0$ can be considered uniform for NEAT, LA and NP, and is questionable for LA+S. NEAT shows the highest values of sensitivity and AUC, and its specificity is exactly equal to the target value (95\%).
}\label{Tab:simd}
	\vspace{0.4cm} \centering
	\begin{tabular}{ccccccc}
		\hline
Test & $p^{KS}$ & $R_1$ & $R_5$ & Sensitivity & Specificity & AUC\\ \hline
NEAT	& \textbf{0.343} & 0.62 		& \textbf{0.98} & \textbf{79}\% & \textbf{95}\% & \textbf{0.925}\\
NEA		& 0.024 		& 0    			& 0.82 			& 73\% 			& 96\% 			& 0.912\\
LP		& 0				& 1.33 			& 1.51 			& \textbf{78}\% & 92\% 			& 0.904\\
LA		& \textbf{0.111} & \textbf{1.16} & 1.33 		& 73\% 			& 93\% 			& 0.908\\
LA+S	& 0.024 		& \textbf{1.16} & 1.13 			& 76\% 			& 94\% 			& 0.910\\
NP		& \textbf{0.323} & 1.42 		& 1.16 			& 70\% 			& 94\% 			& 0.908\\ \hline
	\end{tabular}
\end{table}

\begin{table}[tb]
	\caption{\textbf{Speed comparison.} The table compares the time (in seconds) that each method required to compute 1225 tests for enrichment in simulations S4 and S5, using a processor with 2.5 GhZ CPU frequency. NEAT turns out to be by far the fastest method.
	}\label{Tab:speed_sim}
	\vspace{0.4cm} \centering
	\begin{tabular}{cccc}
		\hline
		 Test & Software & Simulation S4 & Simulation S5\\\hline
		 NEAT 	& \texttt{R} package \texttt{neat} 		& 0.6 & 0.7\\
		 NEA 	& \texttt{R} package \texttt{neaGUI} 	& 2125.4 & 2151.5\\
		 LP 	& CrossTalkZ 							& 28.6 & 44.7\\
		 LA 	& CrossTalkZ 							& 14.4 & 18.0\\
		 LA+S 	& CrossTalkZ 							& 21.8 & 27.6\\
		 NP 	& CrossTalkZ 							& 12.9 & 15.8 \\\hline
	\end{tabular}
\end{table}

\section{Network enrichment analysis: an application to yeast}\label{sec:yeast}
The budding yeast \textit{Saccharomyces cerevisiae} is a unicellular eukaryote organism that can be easily grown in laboratory. Because of these features, it represents a model organism that has been extensively studied, and it was the first eukaryote whose genome was completely sequenced \citep{goffeau1996}. Since then, a large number of studies has aimed to detect associations between genes. In an attempt to collect these results into a unique source, \cite{kim2013} developed \textit{YeastNet}, an undirected gene network that aims to integrate the results of a large number of high-throughput studies on Saccharomyces cerevisiae. In its most recent version (v3), YeastNet comprises 362512 edges connecting 5808 genes. We use this network of known associations in the following analyses.

\subsection{Network enrichment analysis of environmental stress response in yeast}

After analysing gene expression patterns of yeast Saccharomyces cerevisiae in response to different stressful stimuli, \cite{gasch2000} inferred the existence of a set of 868 genes that reacted in a similar way to different, hostile environmental changes. This set of genes, called \textit{Environmental Stress Response} (ESR), is believed to constitute a coordinated, initial reaction to the emergence of any hostile condition in the cell. It consists of two subgroups of genes, containing genes that are repressed and induced under stressful conditions, respectively.\\
We take these two gene sets as target sets, and for each of them we test enrichment with the following functional gene sets: 99 gene sets that are part of the GO Slim biological process ontology (we do not consider the groups ``biological process'' and ``other'' in the analysis) and 106 known KEGG pathways. \\
At $\alpha = 1\%$ level, NEAT detects over-enrichment between 23 GO Slim sets and the set of repressed genes, and between 25 GO Slim sets and the set of induced genes. Furthermore, 15 KEGG pathways are found to be over-enriched with the set of repressed ESR genes, and 47 with the set of induced genes.\\
\cite{gasch2000} reports that genes that are repressed in the ESR are involved in growth related processes, various aspects of RNA metabolism, nucleotide biosyntesis, secretion, encoding of ribosomal proteins and other metabolic processes. These results are in strong agreement with the list of over-enrichments detected by NEAT, shown in Table \ref{esr1}. As a matter of fact, most of the over-enrichments detected by NEAT are related to RNA transcription, nucleotide secretion and translation of ribosomal proteins (rows 1-18 and 24-35 in Table \ref{esr1}), growth-related processes (row 22) and further metabolic processes (rows 23 and 33-35).\\
\cite{gasch2000} observed that inference for the set of genes that are induced by the ESR is more complicated, because most of the genes in this group lack functional annotations. It is worthwhile to observe that NEAT detects a large number of enriched KEGG pathways (47 out of 106). This preliminary observation points out a major feature of the Environmental Stress Response: the cell reacts to the emergence of different hostile conditions by activating a number of known cellular pathways that involve energy production, metabolic reactions and molecular transportation (see Table \ref{esr2b}).\\
Our results for this gene set do not only match the ones of the original study - identifying many processes and pathways that are related to carbohydrate metabolism (rows 1-3 in Table \ref{esr2} and 1-9 in Table \ref{esr2b}), fatty acid metabolism (rows 4-6 in Table \ref{esr2} and 10-18 in Table \ref{esr2b}), mythocondrial functions and cellular redox reactions (rows 5-9 in Table \ref{esr2} and 19-21 in Table \ref{esr2b}), protein folding and degradation (10 in Table \ref{esr2} and 22 in Table \ref{esr2b}) and cellular protection during stressful conditions (rows 11-13 in Table \ref{esr2} and 23 in Table \ref{esr2b}) - but they also unveil further enrichments that involve molecular transportation (rows 3, 6, 14-18 in Table \ref{esr2}) and amino-acid metabolism (rows 24-36 in Table \ref{esr2b}).\\
Tables \ref{cfr1}, \ref{cfr2go} and \ref{cfr2kegg} compare the p-values obtained with NEAT with those obtained with LA+S \citep{mccormack2013}, which, according to the conclusions of \cite{mccormack2013} and to our own simulations, can be considered as the main competitor of NEAT. The tables show a large overlap between the over-enrichments detected by the two methods at a $1\%$ significance level: the two methods jointly detect 34 over-enrichments (19 GO Slim sets and 15 KEGG pathways) for the set of repressed ESR genes, and 67 (24 GO Slim sets and 43 KEGG pathways) for the set of induced ESR genes. There is only a small number of discrepancies between the two methods and these are mostly borderline cases. In particular, LA+S detects 4 over-enrichments that are not detected by NEAT (rows 39 in Table \ref{cfr1}, 26-27 in Table \ref{cfr2go} and 48 in Table \ref{cfr2kegg}), whereas NEAT detects 9 over-enrichments that are not detected by LA+S (rows 19-22 in Table \ref{cfr1}, 25 in Table \ref{cfr2go} and 43-46 in Table \ref{cfr2kegg}).  As concerns computing time, NEAT computed the required task (410 tests in total) in 23 seconds, whereas the same computation with LA+S required 1171 seconds. In summary, the two methods lead to very similar conclusions, but NEAT is considerably more efficient.

\begin{table}[tb]
	\caption{\textbf{Network enrichment analysis of the repressed ESR gene set.} The table lists the 23 Go Slim BP gene sets and the 15 KEGG pathways which the set of repressed ESR genes is found to be over-enriched with at 1\% significance level.}\label{esr1}
\footnotesize \centering
	\begin{tabular}{lllll}
		\hline
  & Gene set & $n_{AB}$ & $\mu_0$ & $log_{10}$(p-value)\\\hline
  & \textbf{Go Slim BP sets:}\\
1 & cytoplasmic translation & 6878 & 2641.9 &    $<$-300 \\
2 & ribosomal large subunit biogenesis & 3408 & 1097.8 &    $<$-300 \\
3 & ribosomal small subunit biogenesis & 5861 & 2073.7 &    $<$-300 \\
4 & ribosome assembly & 1782 & 621.9 &    $<$-300 \\
5 & RNA modification & 2944 & 1062.0 &    $<$-300 \\
6 & rRNA processing & 9187 & 3290.2 &    $<$-300 \\
7 & tRNA processing & 2037 & 901.0 & $<$-300 \\
8 & translational elongation & 1786 & 782.3 & -283.8 \\
9 & ribosomal subunit export from nucleus & 1420 & 561.4 & -281.8 \\
10 & translational initiation & 939 & 462.5 & -112.1 \\
11 & transcription from RNA polymerase III promoter & 565 & 228.4 & -107.7 \\
12 & snoRNA processing & 634 & 303.3 &  -82.0 \\
13 & regulation of translation & 1952 & 1328.6 &  -73.5 \\
14 & DNA-dependent transcription, termination & 774 & 447.0 &  -57.5 \\
15 & transcription from RNA polymerase I promoter & 1005 & 646.4 &  -49.5 \\
16 & protein alkylation & 1063 & 759.4 &  -31.4 \\
17 & tRNA aminoacylation for protein translation & 400 & 233.1 &  -29.4 \\
18 & peptidyl-amino acid modification & 1088 & 883.0 &  -13.2 \\  
  
19 & nuclear transport & 3154 & 2003.5 & -162.4 \\
20 & organelle assembly & 2090 & 1362.7 &  -96.1 \\
21 & nucleobase-containing compound transport & 1453 & 1155.4 &  -20.8 \\
22 & cytokinesis & 1024 & 806.9 &  -16.0 \\
23 & vitamin metabolic process & 325 & 274.0 &   -3.1 \\

    & \textbf{KEGG pathways:}\\
  24 & Ribosome biogenesis in eukaryotes & 9824 & 3661.0 &   $<$-300 \\
  25 & Ribosome & 18640 & 8731.7 &   $<$-300 \\
  26 & RNA polymerase & 3057 & 1541.2 &   $<$-300 \\
  27 & RNA transport & 4341 & 2906.4 & -177.6 \\
  28 & Aminoacyl-tRNA biosynthesis & 1433 & 960.9 &  -58.2 \\
  29 & RNA degradation & 2560 & 1939.3 &  -51.9 \\
  30 & mRNA surveillance pathway & 1768 & 1413.5 &  -24.0 \\
  31 & Pentose phosphate pathway & 1126 & 947.1 &   -9.7 \\
  32 & Spliceosome & 2649 & 2523.6 &   -2.3 \\ 
  33 & Purine metabolism & 5579 & 3623.0 & -263.6 \\
  34 & Pyrimidine metabolism & 4541 & 2884.5 & -234.9 \\
  35 & Cyanoamino acid metabolism & 218 & 158.8 &   -6.3 \\
  36 & One carbon pool by folate & 541 & 392.5 &  -15.0 \\
  37 & Sulfur relay system & 238 & 196.5 &   -2.9 \\
  38 & Carbapenem biosynthesis & 117 & 89.8 &   -2.7 \\  \hline
  	\end{tabular}
\end{table}

\begin{table}[tb]
	\caption{\textbf{Network enrichment analysis of the induced ESR gene set (GO Slim sets).} The table lists the 25 Go Slim BP gene sets which the set of induced ESR genes is found to be over-enriched with at 1\% significance level.}\label{esr2}
\small \centering
	\begin{tabular}{lllll}
		\hline
  & GO Slim BP gene set & $n_{AB}$ & $\mu_0$ & $log_{10}$(p-value)\\\hline
1 & carbohydrate metabolic process & 1296 & 671.2 & -110.9 \\  
2 & oligosaccharide metabolic process & 442 & 165.3 &  -77.3 \\  
3 & carbohydrate transport & 202 & 65.8 &  -45.0 \\

4 & lipid metabolic process & 693 & 484.4 &  -19.9 \\
5 & peroxisome organization & 181 & 124.8 &   -6.0 \\
6 & lipid transport & 120 & 79.7 &   -4.9 \\

7 & generation of precursor metabolites and energy & 585 & 294.8 &  -54.0 \\
8 & cellular respiration & 210 & 118.4 &  -14.5 \\
9 & proteolysis involved in cellular protein catabolic proc. & 639 & 488.5 &  -10.9 \\

10 & protein folding & 476 & 296.9 &  -22.7 \\

11 & response to oxidative stress & 813 & 242.2 & -202.7 \\
12 & response to chemical stimulus & 1489 & 885.1 &  -83.4 \\  
13 & response to starvation & 459 & 331.4 &  -11.2 \\

14 & transmembrane transport & 910 & 644.4 &  -24.2 \\
15 & endocytosis & 395 & 245.5 &  -19.3 \\
16 & protein targeting & 628 & 478.8 &  -10.9 \\
17 & ion transport & 464 & 380.2 &   -4.8 \\
18 & amino acid transport & 137 & 109.4 &   -2.1 \\

19 & cofactor metabolic process & 523 & 219.0 &  -73.7 \\
20 & nucleobase-containing small molecule metabolic proc. & 722 & 404.5 &  -49.2 \\
21 & membrane invagination & 278 & 120.6 &  -37.0 \\
22 & vacuole organization & 335 & 200.2 &  -18.9 \\
23 & protein maturation & 49 & 27.7 &   -3.9 \\
24 & cell morphogenesis & 113 & 79.4 &   -3.6 \\
25 & sporulation & 352 & 306.4 &   -2.1 \\  \hline
  	\end{tabular}
\end{table}

\begin{table}[tb]
	\caption{\textbf{Network enrichment analysis of the induced ESR gene set (KEGG pathways)}.The table lists the 47 KEGG pathways which the set of induced ESR genes is found to be over-enriched with at 1\% significance level.}\label{esr2b}
\scriptsize \centering
	\begin{tabular}{lllll}
		\hline
		& KEGG pathway & $n_{AB}$ & $\mu_0$ & $log_{10}$(p-value)\\\hline
1 & Starch and sucrose metabolism & 1436 & 394.2 &   $<$-300 \\
2 & Pentose and glucuronate interconversions & 414 & 110.7 & -119.9 \\
3 & Glycolysis / Gluconeogenesis & 1235 & 616.3 & -116.5 \\
4 & Fructose and mannose metabolism & 562 & 200.0 & -106.7 \\
5 & Galactose metabolism & 511 & 173.9 & -104.5 \\
6 & Amino sugar and nucleotide sugar metabolism & 567 & 264.2 &  -63.4 \\
7 & Other glycan degradation & 79 & 11.7 &  -44.2 \\
8 & Pyruvate metabolism & 633 & 355.9 &  -42.8 \\
9 & Propanoate metabolism & 189 & 107.3 &  -12.9 \\

10 & Glycerolipid metabolism & 444 & 172.1 &  -72.7 \\
11 & Peroxisome & 633 & 313.3 &  -61.2 \\
12 & Fatty acid degradation & 419 & 215.0 &  -37.2 \\
13 & Arachidonic acid metabolism & 117 & 36.7 &  -28.1 \\
14 & Sphingolipid metabolism & 227 & 103.6 &  -27.3 \\
15 & Glycerophospholipid metabolism & 450 & 270.9 &  -24.5 \\
16 & alpha-Linolenic acid metabolism & 69 & 27.1 &  -11.7 \\
17 & Fatty acid elongation & 138 & 75.3 &  -10.8 \\
18 & Biosynthesis of unsaturated fatty acids & 134 & 103.9 &   -2.5 \\

19 & Glutathione metabolism & 467 & 204.8 &  -59.9 \\
20 & Citrate cycle (TCA cycle) & 487 & 267.3 &  -35.6 \\
21 & Ubiquinone and other terpenoid-quinone biosynthesis & 96 & 41.8 &  -13.1 \\

22 & Protein processing in endoplasmic reticulum & 1121 & 866.0 &  -17.4 \\

23 & Longevity regulating pathway & 987 & 544.0 &  -70.6 \\

24 & beta-Alanine metabolism & 397 & 104.0 & -118.0 \\
25 & Taurine and hypotaurine metabolism & 132 & 24.3 &  -59.4 \\
26 & Tyrosine metabolism & 382 & 163.5 &  -51.8 \\
27 & Tryptophan metabolism & 292 & 113.3 &  -48.2 \\
28 & Valine, leucine and isoleucine degradation & 276 & 107.5 &  -45.3 \\
29 & Alanine, aspartate and glutamate metabolism & 488 & 262.2 &  -38.0 \\
30 & Histidine metabolism & 267 & 127.4 &  -28.8 \\
31 & Arginine and proline metabolism & 301 & 154.3 &  -27.0 \\
32 & Lysine degradation & 294 & 150.4 &  -26.6 \\
33 & Phenylalanine metabolism & 171 & 71.4 &  -25.0 \\
34 & Glycine, serine and threonine metabolism & 350 & 264.3 &   -6.7 \\
35 & Cysteine and methionine metabolism & 338 & 285.3 &   -2.8 \\
36 & Arginine biosynthesis & 167 & 134.0 &   -2.4 \\

37 & Butanoate metabolism & 460 & 84.8 & -202.8 \\
38 & Pentose phosphate pathway & 604 & 288.0 &  -64.0 \\
39 & Regulation of autophagy & 303 & 126.7 &  -43.3 \\
40 & Insulin resistance & 337 & 172.8 &  -30.1 \\
41 & Glyoxylate and dicarboxylate metabolism & 368 & 201.6 &  -27.3 \\
42 & Methane metabolism & 435 & 254.2 &  -26.2 \\
43 & Nicotinate and nicotinamide metabolism & 154 & 99.8 &   -6.7 \\
44 & Nitrogen metabolism & 88 & 52.8 &   -5.4 \\
45 & Thiamine metabolism & 57 & 32.9 &   -4.1 \\
46 & Selenocompound metabolism & 122 & 89.3 &   -3.2 \\
47 & Sulfur metabolism & 133 & 105.3 &   -2.2 \\ 	\hline
\end{tabular}
\end{table}	

\begin{table}[tb]
	\caption{\textbf{Repressed ESR gene set: comparison between NEAT and LA+S.} The table reports the gene sets that are found to be over-enriched ($\alpha = 1 \%$) by at least one of the two methods. $\mu_0$ denotes the expected value of $N_{AB}$ in the absence of enrichment. The last two columns report $log_{10}$ p-values for the proposed NEAT and the LA+S test of \cite{mccormack2013}, respectively.}\label{cfr1}
\scriptsize \centering
	\begin{tabular}{llllll}
		\hline
		& & \multicolumn{2}{c}{$\mu_0$} & \multicolumn{2}{c}{$\log_{10}$(p-value)}\\
 & Gene set & NEAT & LA+S & NEAT & LA+S\\\hline
 &  \textbf{GO Slim BP sets:}\\
1 & cytoplasmic translation & 2641.9 & 3583.5 & $<$-300 & -290.9 \\ 
2 & ribosomal large subunit biogenesis & 1097.8 & 1602.4 & $<$-300 & -269.2 \\ 
3 & ribosomal small subunit biogenesis & 2073.7 & 3013.2 & $<$-300 & -236.8 \\ 
4 & ribosome assembly & 621.9 & 872.1 & $<$-300 & -95.9 \\ 
5 & RNA modification & 1062.0 & 1422.7 & $<$-300 & -213.7 \\ 
6 & rRNA processing & 3290.2 & 4623.2 & $<$-300 & $<$-300 \\ 
7 & tRNA processing & 901.0 & 1137.6 & $<$-300 & -103.3 \\ 
8 & translational elongation & 782.3 & 1019.5 & -283.8 & -71.2 \\ 
9 & ribosomal subunit export from nucleus & 561.4 & 693.4 & -281.8 & -151.2 \\ 
10 & nuclear transport & 2003.5 & 2452.5 & -162.4 & -33.0 \\ 
11 & translational initiation & 462.5 & 594.8 & -112.1 & -33.6 \\ 
12 & transcription from RNA polymerase III promoter & 228.4 & 281.6 & -107.7 & -43.6 \\ 
13 & organelle assembly & 1362.7 & 1719.2 & -96.1 & -8.0 \\ 
14 & snoRNA processing & 303.3 & 349.8 & -82.0 & -26.5 \\ 
15 & regulation of translation & 1328.6 & 1577.5 & -73.5 & -12.9 \\ 
16 & DNA-dependent transcription, termination & 447.0 & 575.2 & -57.5 & -11.7 \\ 
17 & transcription from RNA polymerase I promoter & 646.4 & 874.2 & -49.5 & -5.2 \\ 
18 & tRNA aminoacylation for protein translation & 233.1 & 256.7 & -29.4 & -11.2 \\ 
19 & protein alkylation & 759.4 & 1000.0 & -31.4 & -1.2 \\ 
20 & nucleobase-containing compound transport & 1155.4 & 1445.1 & -20.8 & -0.1 \\ 
21 & cytokinesis & 806.9 & 925.9 & -16.0 & -1.8 \\ 
22 & peptidyl-amino acid modification & 883.0 & 1102.4 & -13.2 & -0.1 \\ 
23 & vitamin metabolic process & 274.0 & 245.8 & -3.1 & -5.5 \\ 
  & \textbf{KEGG pathways:}\\
24 & Ribosome biogenesis in eukaryotes & 3661.0 & 5212.5 &   $<$-300 &   $<$-300 \\ 
25 & Ribosome & 8731.7 & 11954.0 &   $<$-300 & -283.3 \\ 
26 & RNA polymerase & 1541.2 & 2058.0 &   $<$-300 &  -76.1 \\ 
27 & Purine metabolism & 3623.0 & 4136.9 & -263.6 &  -66.9 \\ 
28 & Pyrimidine metabolism & 2884.5 & 3402.5 & -234.9 &  -61.0 \\ 
29 & RNA transport & 2906.4 & 3193.2 & -177.6 &  -75.4 \\ 
30 & Aminoacyl-tRNA biosynthesis & 960.9 & 934.2 &  -58.2 &  -49.8 \\ 
31 & RNA degradation & 1939.3 & 2051.3 &  -51.9 &  -19.9 \\ 
32 & mRNA surveillance pathway & 1413.5 & 1477.3 &  -24.0 &  -12.7 \\ 
33 & One carbon pool by folate & 392.5 & 344.2 &  -15.0 &  -19.5 \\ 
34 & Pentose phosphate pathway & 947.1 & 979.2 &   -9.7 &   -4.6 \\ 
35 & Cyanoamino acid metabolism & 158.8 & 132.2 &   -6.3 &   -7.2 \\ 
36 & Sulfur relay system & 196.5 & 172.7 &   -2.9 &   -3.9 \\ 
37 & Carbapenem biosynthesis & 89.8 & 75.1 &   -2.7 &   -4.1 \\ 
38 & Spliceosome & 2523.6 & 2432.2 &   -2.3 &   -4.1 \\ 
39 & Synthesis and degradation of ketone bodies & 39.8 & 29.8 &   -0.3 &   -2.2 \\ \hline 
  	\end{tabular}
\end{table}

\begin{table}[tb]
	\caption{\textbf{Induced ESR gene set: comparison between NEAT and LA+S (GO Slim sets)}. The table reports the gene sets that are found to be over-enriched ($\alpha = 1 \%$) by at least one of the two methods. $\mu_0$ denotes the expected value of $N_{AB}$ in the absence of enrichment. The last two columns report $log_{10}$ p-values for the proposed NEAT and the LA+S test of \cite{mccormack2013}, respectively.}\label{cfr2go}
\footnotesize \centering
	\begin{tabular}{llllll}
		\hline
		& & \multicolumn{2}{c}{$\mu_0$} & \multicolumn{2}{c}{$\log_{10}$(p-value)}\\
 & GO Slim BP set & NEAT & LA+S & NEAT & LA+S\\\hline
1 & response to oxidative stress & 242.2 & 248.5 & -202.7 & -253.7 \\ 
2 & carbohydrate metabolic process & 671.2 & 663.9 & -110.9 & -123.3 \\ 
3 & response to chemical stimulus & 885.1 & 912.4 & -83.4 & -92.8 \\ 
4 & oligosaccharide metabolic process & 165.3 & 158.1 & -77.3 & -104.5 \\ 
5 & cofactor metabolic process & 219.0 & 225.6 & -73.7 & -76.2 \\ 
6 & generation of precursor metabolites and energy & 294.8 & 293.4 & -54.0 & -56.1 \\ 
7 & nucleobase-containing small molecule metabolic process & 404.5 & 417.4 & -49.2 & -41.0 \\ 
8 & carbohydrate transport & 65.8 & 77.7 & -45.0 & -52.8 \\ 
9 & membrane invagination & 120.6 & 118.3 & -37.0 & -51.7 \\ 
10 & transmembrane transport & 644.4 & 684.7 & -24.2 & -16.2 \\ 
11 & protein folding & 296.9 & 296.3 & -22.7 & -26.6 \\ 
12 & lipid metabolic process & 484.4 & 495.7 & -19.9 & -23.3 \\ 
13 & endocytosis & 245.5 & 248.7 & -19.3 & -19.3 \\ 
14 & vacuole organization & 200.2 & 199.7 & -18.9 & -22.4 \\ 
15 & cellular respiration & 118.4 & 125.2 & -14.5 & -14.1 \\ 
16 & response to starvation & 331.4 & 318.4 & -11.2 & -15.8 \\ 
17 & protein targeting & 478.8 & 485.1 & -10.9 & -15.8 \\ 
18 & proteolysis involved in cellular protein catabolic process & 488.5 & 494.1 & -10.9 & -9.8 \\ 
19 & peroxisome organization & 124.8 & 123.5 & -6.0 & -6.0 \\ 
20 & lipid transport & 79.7 & 90.4 & -4.9 & -2.8 \\ 
21 & ion transport & 380.2 & 410.7 & -4.8 & -2.1 \\ 
22 & protein maturation & 27.7 & 30.9 & -3.9 & -3.0 \\ 
23 & cell morphogenesis & 79.4 & 80.8 & -3.6 & -3.7 \\ 
24 & sporulation & 306.4 & 301.7 & -2.1 & -2.5 \\ 
25 & amino acid transport & 109.4 & 113.0 & -2.1 & -1.6 \\ 
26 & response to osmotic stress & 181.8 & 178.3 & -1.6 & -2.1 \\ 
27 & protein phosphorylation & 587.6 & 564.3 & -1.4 & -2.7 \\ 
\hline
  	\end{tabular}
\end{table}

\begin{table}[tb]
	\caption{\textbf{Induced ESR gene set: comparison between NEAT and LA+S (KEGG pathways)}.
	%The table reports the gene sets that are found to be over-enriched ($\alpha = 1 \%$) by at least one of the two methods. $\mu_0$ denotes the expected value of $N_{AB}$ in absence of enrichment. The last two columns report $log_{10}$ p-values for the proposed NEAT and the LA+S test of \cite{mccormack2013}, respectively.
	}\label{cfr2kegg}
\scriptsize \centering
	\begin{tabular}{llllll}
		\hline
		& & \multicolumn{2}{c}{$\mu_0$} & \multicolumn{2}{c}{$\log_{10}$(p-value)}\\
 & KEGG pathway & NEAT & LA+S & NEAT & LA+S\\\hline
1 & Starch and sucrose metabolism & 394.2 & 400.6 &   $<$-300 &   $<$-300 \\ 
2 & Butanoate metabolism & 84.8 & 98.0 & -202.8 &   $<$-300 \\ 
3 & Pentose and glucuronate interconversions & 110.7 & 127.5 & -119.9 & -185.7 \\ 
4 & beta-Alanine metabolism & 104.0 & 122.9 & -118.0 & -209.8 \\ 
5 & Glycolysis / Gluconeogenesis & 616.3 & 618.7 & -116.5 & -149.3 \\ 
6 & Fructose and mannose metabolism & 200.0 & 206.2 & -106.7 & -160.7 \\ 
7 & Galactose metabolism & 173.9 & 193.2 & -104.5 & -126.4 \\ 
8 & Glycerolipid metabolism & 172.1 & 193.2 &  -72.7 & -103.2 \\ 
9 & Longevity regulating pathway - multiple species & 544.0 & 508.2 &  -70.6 &  -79.1 \\ 
10 & Pentose phosphate pathway & 288.0 & 284.2 &  -64.0 & -105.8 \\ 
11 & Amino sugar and nucleotide sugar metabolism & 264.2 & 277.6 &  -63.4 &  -66.7 \\ 
12 & Peroxisome & 313.3 & 332.9 &  -61.2 &  -55.8 \\ 
13 & Glutathione metabolism & 204.8 & 221.6 &  -59.9 &  -77.8 \\ 
14 & Taurine and hypotaurine metabolism & 24.3 & 28.5 &  -59.4 &  -92.8 \\ 
15 & Tyrosine metabolism & 163.5 & 169.9 &  -51.8 &  -62.6 \\ 
16 & Tryptophan metabolism & 113.3 & 130.9 &  -48.2 &  -59.4 \\ 
17 & Valine, leucine and isoleucine degradation & 107.5 & 124.8 &  -45.3 &  -56.8 \\ 
18 & Other glycan degradation & 11.7 & 12.9 &  -44.2 &  -66.3 \\ 
19 & Regulation of autophagy & 126.7 & 135.2 &  -43.3 &  -45.5 \\ 
20 & Pyruvate metabolism & 355.9 & 388.8 &  -42.8 &  -41.6 \\ 
21 & Alanine, aspartate and glutamate metabolism & 262.2 & 284.5 &  -38.0 &  -36.7 \\ 
22 & Fatty acid degradation & 215.0 & 225.0 &  -37.2 &  -43.7 \\ 
23 & Citrate cycle (TCA cycle) & 267.3 & 299.5 &  -35.6 &  -32.9 \\ 
24 & Insulin resistance & 172.8 & 176.5 &  -30.1 &  -30.4 \\ 
25 & Histidine metabolism & 127.4 & 147.8 &  -28.8 &  -25.8 \\ 
26 & Arachidonic acid metabolism & 36.7 & 44.1 &  -28.1 &  -40.6 \\ 
27 & Glyoxylate and dicarboxylate metabolism & 201.6 & 224.8 &  -27.3 &  -23.7 \\ 
28 & Sphingolipid metabolism & 103.6 & 116.3 &  -27.3 &  -26.2 \\ 
29 & Arginine and proline metabolism & 154.3 & 180.2 &  -27.0 &  -24.8 \\ 
30 & Lysine degradation & 150.4 & 160.2 &  -26.6 &  -31.5 \\ 
31 & Methane metabolism & 254.2 & 262.7 &  -26.2 &  -23.7 \\ 
32 & Phenylalanine metabolism & 71.4 & 81.5 &  -25.0 &  -26.4 \\ 
33 & Glycerophospholipid metabolism & 270.9 & 285.1 &  -24.5 &  -22.3 \\ 
34 & Protein processing in endoplasmic reticulum & 866.0 & 857.1 &  -17.4 &  -20.7 \\ 
35 & Ubiquinone and other terpenoid-quinone biosynthesis & 41.8 & 47.1 &  -13.1 &  -12.3 \\ 
36 & Propanoate metabolism & 107.3 & 122.9 &  -12.9 &   -9.9 \\ 
37 & alpha-Linolenic acid metabolism & 27.1 & 30.5 &  -11.7 &  -11.2 \\ 
38 & Fatty acid elongation & 75.3 & 76.1 &  -10.8 &  -12.9 \\ 
39 & Glycine, serine and threonine metabolism & 264.3 & 281.1 &   -6.7 &   -3.5 \\ 
40 & Nicotinate and nicotinamide metabolism & 99.8 & 111.9 &   -6.7 &   -4.7 \\ 
41 & Nitrogen metabolism & 52.8 & 60.7 &   -5.4 &   -4.0 \\ 
42 & Thiamine metabolism & 32.9 & 36.8 &   -4.1 &   -3.2 \\ 
43 & Selenocompound metabolism & 89.3 & 97.0 &   -3.2 &   -1.9 \\ 
44 & Cysteine and methionine metabolism & 285.3 & 310.6 &   -2.8 &   -1.0 \\ 
45 & Arginine biosynthesis & 134.0 & 154.2 &   -2.4 &   -0.6 \\ 
46 & Sulfur metabolism & 105.3 & 121.9 &   -2.2 &   -0.5 \\ 
47 & Biosynthesis of unsaturated fatty acids & 103.9 & 102.1 &   -2.5 &   -3.1 \\ 
48 & Regulation of mitophagy - yeast & 554.4 & 510.4 &   -1.6 &   -5.1 \\ 
\hline
\end{tabular}
\end{table}	

\subsection{Network enrichment analysis of GO Slim sets: overlap does not imply enrichment}
\begin{figure}[tb]
\begin{center}
	\includegraphics[scale=0.5]{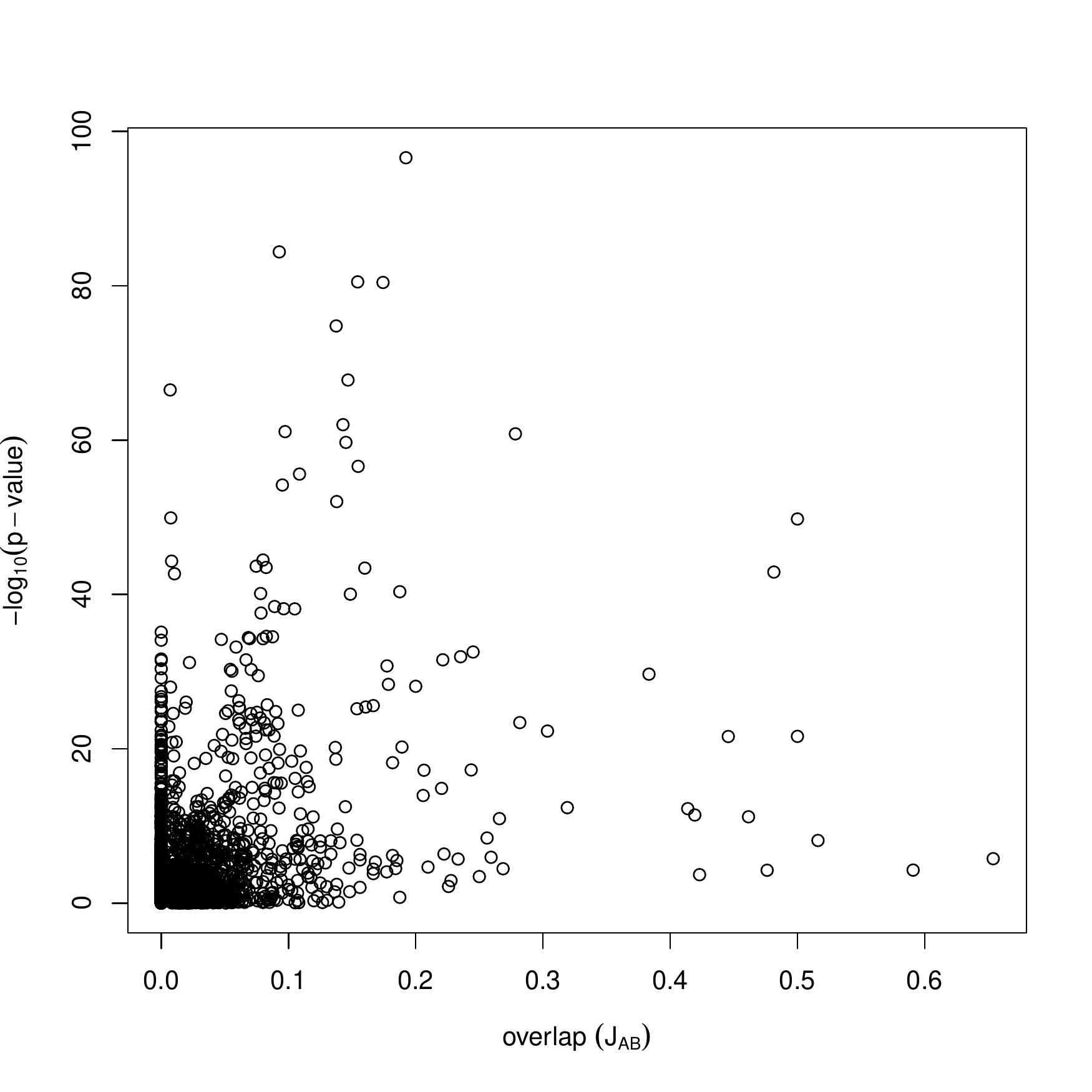}
\end{center}
	\caption{\textbf{Relation between overlap ($J_{AB}$) and p-values.}
	Note that p-values are represented on a negative log-scale to enhance readability.
	}\label{fig:pval-overlap}
\end{figure}

Gene ontologies \citep{ashburner2000} consist of a large number of gene sets, which are involved in different cellular functions or biological processes, or that are active in a specific component of the cell. These sets of genes are typically employed to enrich sets of differentially expressed genes that have been experimentally detected (the analysis of the ESR gene sets in the previous subsection provides an example of this). However, network enrichment analysis is a more general instrument, which allows to assess the relation between pairs of gene sets in a network. 
One might wonder, for instance, whether gene sets within an ontology tend to be strongly related to each other, or whether there is a strong separation between them.\\
We consider gene sets in the GO Slim biological process ontology for \textit{Saccharomyces cerevisiae} (we once more exclude the two general groups ``biological process'' and ``other'' from the analysis). As a result of the hierarchical structure of Gene Ontologies, 12 gene sets are nested within another group. We exclude these 12 sets from the analysis: the remaining 87 gene sets do not have hierarchical relations with each other, and pairs of these sets display overall a low overlap (1.7 \% on average), which is null in most cases (62\% of pairs of sets do not share genes). If overlapping of sets was taken by itself as evidence of a relation between two gene sets, one would therefore conclude that most of these gene sets are unrelated.\\
If, however, we do not limit our attention to the overlap between pairs of sets, but consider also known associations between genes in the two sets as represented in YeastNet \citep{kim2013}, we obtain a different conclusion. We have used NEAT to test whether there is enrichment between each pair of sets. In a random network where no relations between the sets are present, we would expect to detect $37$ enrichments (on average) out of $3741$ tests for $\alpha = 1\%$; instead, we detect $1409$ enrichments, $38$ times more than expected. Out of these, $710$ are under-enrichments, and $699$ are over-enrichments. An under-enrichment, here, indicates that two GO Slim sets are poorly connected to each other: the high number of under-enrichments, therefore, might be not particularly surprising or interesting, as we do expect that unrelated gene sets within the ontology are poorly connected. The high number of over-enrichments, on the other hand, is striking: this indicates that many groups within the ontology are highly connected to each other - something that would occur rather rarely, if there was no relation between the sets.\\
This result points out a major difference between gene enrichment analysis and network enrichment analysis: whereas in the first case the extent of overlapping between two gene sets is taken by itself as evidence of enrichment, network enrichment analysis bases the evaluation of enrichment on the level of connectivity that exists between the two sets in a network. Of course, the two facts are not completely unrelated. Figure \ref{fig:pval-overlap} shows that there is a certain correlation between overlap of gene sets (Jaccard index) and network enrichment, so that we tend to find network enrichment in the presence of higher levels of overlap. This correlation is, however, low (the Pearson correlation coefficient between $J_{AB}$ and $p^{AB}$ is -0.15), pointing out that there does not necessarily have to be enrichment for highly overlapping gene sets, and vice versa. As an example, the GO Slim sets ``cytokinesis'' and ``nuclear organization'' do not share genes, but are detected as enriched ($p=0.0003$) in YeastNet. This result can be explained by the fact that ``nuclear organization'' includes genes involved in the assembly and disassembly of the nucleus, which is a preliminary step in cell cytokinesis.

\section{Conclusion}
Network enrichment analysis is a powerful extension of traditional methods of gene enrichment analysis, that allows to integrate them with the information on connectivity between genes provided by genetic networks. Whereas gene enrichment analysis bases the test for enrichment solely on the overlap between two gene sets and ignores the relationships between individual genes, network enrichment analysis exploits a larger amount of information by making use of gene networks, and it is thus capable to detect enrichment even between two gene sets that do not share genes.\\
In this paper, we have presented a Network Enrichment Analysis Test (NEAT) that aims to overcome some limitations which affect the network enrichment tests of \cite{alexeyenko2012} and \cite{mccormack2013}. First of all, we believe that a normal approximation does not make justice to the discrete nature of $N_{AB}$. We have shown that this approximation can be avoided if one models $N_{AB}$ directly, using a hypergeometric distribution with suitably specified parameters. In addition, the normal approximation employed by \cite{alexeyenko2012} and \cite{mccormack2013} requires the computation of a large number of network permutations to obtain the mean and variance under $H_0$: this operation can be very time consuming for big networks and it makes the computation of the test rather slow. The use of the hypergeometric distribution, instead, allows to specify the null distribution of $N_{AB}$ without resorting to permutations, thus speeding up computations considerably. A further drawback of existing methods for network enrichment analysis \citep{shojaie2010,glaab2012,alexeyenko2012,mccormack2013} is that they have been implemented only for undirected networks. We address this problem by considering different types of networks (directed, undirected and partially directed) and by proposing two different parametrizations, which take into account the different nature of directed and undirected links.\\
We believe that NEAT could constitute a flexible and computationally efficient test for network enrichment analysis. Our simulations show that NEAT has a good capacity to correctly classify enrichments and non-enrichments. Comparison of NEAT with other methods points out an overall good performance in terms of sensitivity and of specificity, as well as the computational efficiency of the proposed method. The examples illustrated in the previous Section show that NEAT can retrieve enrichments that were detected with gene enrichment analysis, but it can also unveil further enrichments that would be overlooked, if known associations between genes were ignored. Even though the focus of this paper is on gene regulatory networks, NEAT is a rather general test: it can be applied to networks that arise in different contexts and disciplines, whenever the interest is to infer the relationship between groups of vertices. This can include, for example, other types of biological networks, as well as social, economic or technological networks.

\bibliographystyle{apa}
\bibliography{bibliography-v3}

\end{document}